\documentclass[10pt]{revtex4}
\usepackage{amssymb}
\usepackage{latexsym}
\usepackage{epsfig}

\begin{document}
\title{The upper critical magnetic field of holographic
superconductor with conformally invariant power-Maxwell electrodynamics}
\author{A.\ Sheykhi$^{1,2}$\footnote{asheykhi@shirazu.ac.ir}, F.\ Shamsi$^{1}$,
 and S.\ Davatolhagh$^1$\footnote{davatolhagh@susc.ac.ir}}
\address{$^1$ Physics Department and Biruni Observatory, College of
Sciences, Shiraz University, Shiraz 71946, Iran\\
$^2$ Research Institute for Astronomy and Astrophysics of Maragha
(RIAAM), P.O. Box 55134-441, Maragha, Iran}

\begin{abstract}
The properties of $(d-1)$-dimensional $s$-wave holographic
superconductor in the presence of power-Maxwell field is explored.
We study the probe limit in which the scalar and gauge fields do
not backreact on the background geometry. Our study is based on
the matching of solutions on the boundary and on the horizon at
some intermediate point. At first, the case without external
magnetic field is considered, and the critical temperature is
obtained in terms of the charge density, the dimensionality, and
the power-Maxwell exponent. Then, a magnetic field is turned on
in the $d$-dimensional bulk which can influence the
$(d-1)$-dimensional holographic superconductor at the boundary.
The phase behavior of the corresponding holographic superconductor
is obtained by computing the upper critical magnetic field in the
presence of power-Maxwell
electrodynamics, characterized by the power exponent $q$.
Interestingly, it is observed that in the presence of
magnetic field, the physically acceptable phase behavior of the holographic
superconductor is obtained for $q={d}/{4}$, which guaranties
the conformal invariance of the power-Maxwell Lagrangian. The
case of physical interest in five spacetime dimensions ($d=5$, and $q=5/4$) is considered
in detail, and compared with the results obtained for the usual Maxwell electrodynamics
$q=1$ in the same dimensions.

\end{abstract}

\maketitle
\section{Introduction}
The correspondence between the gravity in a $d$-dimensional
anti-de Sitter (AdS) spacetime and the conformal field theory
(CFT) residing on the ($d-1$)-dimensional boundary of this
spacetime, provides a powerful tool for studying strongly coupled
systems \cite{maldacena}. The gauge/gravity duality, which relates
strongly interacting gauge theories to theories of gravity in
higher dimensions, has opened a new window to study many different
strongly interacting condensed matter systems \cite{Hart}. In
Ref.\ \cite{hartnol}, a model for dual description of a
holographic superconductor was proposed. The model was shown to
exhibit a critical point $T_c$ at which the system goes into a
superconducting phase. The properties of this phase have been
thoroughly studied \cite{Hart,hartnol}, showing a strong resemblance with
those of a Type II superconductor. Till date, a number of attempts
have been made, mostly  numerical, in order to understand various
properties of holographic superconductors in the framework of the
usual Maxwell electromagnetic theory
\cite{hartnol,G,gregory,S.A,zeng,H,cai,Q,rg,Horowitz,How,Pan,Cai1,Pan2}.

It is important to investigate the issue of response to an external
magnetic field in the context of holographic superconductors
\cite{Albash}, which also is of central significance in the general field of superconductivity.
It is observed that when immersed in an external magnetic
field, ordinary superconductors expel the magnetic flux lines
thereby exhibiting perfect diamagnetism when the temperature is
lowered through $ T_{c} $, which is called the Meissner effect
\cite{Tinkham}. In fact depending on their behavior in the
presence of an external magnetic field, the superconductors are
classified into two general categories, namely type I and type II superconductors.
In type I superconductors, when the external magnetic field $B$
reaches a critical value $B_c $ there occurs an abrupt (first
order) phase transition from the superconducting phase to the
normal phase. On the other hand, in type II superconductors that
are directly relevant to our discussion, there happens to be a
continuous (second order) phase transition and the material ceases
to superconduct for $ B > B_{\rm c2} $, where $ B_{\rm c2}$ is
called the upper critical field of type II superconductor. In the
case of the continuous phase transition in type II superconductor,
the order parameter is small near $B_{\rm c2}$ and vanishes
continuously as $B\rightarrow B_{\rm c2}$.

For the holographic superconductor in the probe limit, we neglect
the backreaction of the scalar field on the background geometry.
As a result, the holographic superconductor is not able to repel
the background magnetic field, which holds a strong resemblance
with type II superconductor in its mixed phase. Instead the scalar
condensate adjusts itself such that it only fills a finite strip
in the plane, thus reducing the total magnetic field passing
through it. In other words, the effect of the external magnetic
field is such that it always tries to reduce the condensate. The
numerical studies indicate that the superconducting phase
disappears above an upper critical value of the applied magnetic
field $B > B_{\rm c2} $ \cite{Albash,Nakano}.

It is well known that the properties of holographic superconductor
depend on the behavior of the electromagnetic field coupled with
the charged scalar field. The effects of nonlinear electrodynamics
on the critical temperature and condensation parameters of
$s$-wave holographic superconductors have been investigated  in
\cite{Shey,Zh,GanBI}. Motivated by the recent studies
\cite{Albash2,Roy1,Gan,Xue,Lala,SS,Roy2}, and the fact that within
the framework of AdS/CFT correspondence a different
electromagnetic action is expected to modify the dynamics of the
dual theory, in this paper we investigate the behavior of
holographic superconductor with power-Maxwell field in the
background of a $d$-dimensional Schwarzschild AdS black hole,
bearing in mind that in $ d$-dimensional spacetime the
energy-momentum tensor of conformally invariant Maxwell field is
traceless provided we take $q={d}/{4} $ where $q$ is the power
parameter of the power-Maxwell electrodynamics
\cite{PowerM,SheyPM}. In the absence of external magnetic field,
and in the background of $d$-dimensional AdS black hole, the
properties of $s$-wave holographic superconductors coupled to
 power-Maxwell electrodynamics have been explored in \cite{SSM,PM,PMb}.

The rest of this paper is organized as follows. In Sec.\ II, we
present the holographic dual of $d$-dimensional Schwarzschild AdS
black hole by introducing a complex charged scalar field coupled
with the power Maxwell field and explore the relation between
critical temperature and charge density. In Sec.\ III, the
magnetic field effect on holographic superconductor with power
Maxwell electrodynamics is considered. The paper is concluded with
a discussion and summary in section IV.

\section{Holographic superconductor with power Maxwell field}
We consider the $d$-dimensional action of Einstein gravity in the
presence of power-Maxwell field and a charged complex scalar field
which is given by
\begin{eqnarray}\label{metric}
S=\int
d^{d}x\sqrt{-g}\bigg[R-2\Lambda-\beta(F_{\mu\nu}F^{\mu\nu})^q-
|\nabla_{\mu}\psi - i q A_{\mu} \psi|^{2} - m^{2}|\psi|^{2}\bigg],
\end{eqnarray}
where $R$ is the Ricci scalar, $F_{\mu\nu}$ is the electromagnetic
field tensor, $ A_{\mu} $ and $ \psi $ are the gauge and scalar
field, $ \beta $ is the coupling constant, $ q$ is the power
parameter of the power-Maxwell field, and the cosmological
constant is
\begin{equation}
\Lambda=-\frac{(d-1)(d-2)}{2 l^2},
\end{equation}
where $l$ is the AdS  radius of spacetime. For $ \beta = 1/4 $ and
$ q = 1 $ the power-Maxwell Lagrangian
$\mathcal{L}=-\beta(F_{\mu\nu}F^{\mu\nu})^q$ reduces to the usual
Maxwell Lagrangian. Besides, for $q=d/4$, the energy-momentum
tensor of the power-Maxwell Lagrangian is traceless in all
dimensions \cite{PowerM}. It is also easy to check that, under a
conformal transformation which acts on the fields as
$g_{\mu\nu}\rightarrow \Omega^2 g_{\mu\nu}$ and
$A_{\mu}\rightarrow A_{\mu}$, the power-Maxwell Lagrangian
$\mathcal{L}=-\beta(F_{\mu\nu}F^{\mu\nu})^q$ remains unchanged
provided  $q=d/4$ \cite{PowerM}. The metric of a planar
Schwarzschild AdS black hole in $d$-dimensional spacetime is
\begin{eqnarray}
ds^2=-f(r)dt^2+\frac{dr^2}{f(r)}+r^2dx_{i}dx^{i},
\end{eqnarray}
where
\begin{equation}
f(r)=r^2\bigg(1-\frac{r_{+}^{d-1}}{r^{d-1}}\bigg),
\end{equation}
$r_{+} $ is the horizon radius, and we have taken the AdS
radius equal to one, i.e.\ $ l = 1 $. The Hawking temperature is
given by
\begin{equation}\label{T}
T=\frac{f'(r_+)}{4\pi}=\frac{(d-1)r_{+}}{4\pi}.
\end{equation}
We consider the following ansatz for the gauge and the scalar
fields, respectively \cite{hartnol}
\begin{eqnarray}
A_{\mu}=\phi(r)dt,  \quad  \quad \psi=\psi(r).
\end{eqnarray}
From the action in Eq.\ (\ref{metric}), the equations of motion are given
by
\begin{eqnarray}\label{phir}
\phi''+\bigg(\frac{d-2}{2q-1}\bigg)\frac{\phi'}{r}+\frac{\phi
\psi^2 \phi'^{2-2q}}{(-2)^{q}\beta q(2q-1) f}=0,
\end{eqnarray}
\begin{eqnarray}\label{psir}
\psi''+\bigg(\frac{f'}{f}+\frac{d-2}{r}\bigg)\psi'+\bigg(\frac{\phi^2}{f^2}-\frac{m^2}{f}\bigg)\psi=0,
\end{eqnarray}
where the prime denotes derivative with respect to $r$. At the
event horizon of the black hole $ r = r_{+} $, the regularity
gives the boundary conditions for $ \psi(r) $ and $ \phi(r) $ as
\cite{hartnol}
\begin{equation}\label{ph}
\phi(r_{+})=0, \   \  \   \
 \    \psi(r_{+})=\frac{(d-1)r_{+}}{m^2}\psi'(r_{+}).
\end{equation}
At the asymptotic AdS boundary ($ r\rightarrow \infty  $), the
solutions of Eqs. (\ref{phir})  and (\ref{psir}) behave like
\begin{equation}
\psi\approx\frac{\psi_{-}}{r^{\lambda_{-}}}+\frac{\psi_{+}}{r^{\lambda_{+}}},
\end{equation}
\begin{equation}
\phi\approx\mu-\frac{\rho^{\frac{1}{2q-1}}}{r^{\frac{d-2}{2q-1}-1}},
\end{equation}
with
\begin{eqnarray}
\lambda_{\pm}=\frac{1}{2}\bigg[(d-1)\pm\sqrt{(d-1)^2+4m^2}\bigg],
\end{eqnarray}
where $ \mu $ and $ \rho $ are interpreted as the chemical
potential and charge density in the dual field theory,
respectively \cite{hartnol}. By changing the variable $
z={r_{+}}/{r}$, the equations of motion (\ref{phir}) and
(\ref{psir}) become
\begin{eqnarray}\label{phiz}
\phi''+\left(\frac{4q-d}{2q-1}\right)\frac{\phi'}{z}+\frac{\phi
\psi^2
  \phi'^{2-2q} r_{+}^{2q} }{2^{q} (-1)^{3q}\beta q (2q-1)f z^{4q}}=0,
\end{eqnarray}
\begin{eqnarray}\label{psiz}
\psi''+\left(\frac{f'}{f}-\frac{d-4}{z}\right)\psi'+\frac{r_{+}^2}{z^4}\bigg(\frac{\phi^2}{f^2}-\frac{m^2}{f}\bigg)\psi=0,
\end{eqnarray}
where the prime now indicates the derivative with respect to the
new coordinate $z$. The asymptotic boundary conditions for the
scalar field $ \psi(z) $ and the scalar potential $ \phi(z) $ now
become
\begin{equation}\label{m}
\phi=\mu-\frac{\rho^{\frac{1}{2q-1}}}{r_{+}^{\frac{d-2}{2q-1}-1}}z^{\frac{d-2}{2q-1}-1},
\end{equation}
\begin{equation}\label{J}
\psi=J_{-}z^{\lambda_{-}}+J_{+}z^{\lambda_{+}}.
\end{equation}
Following \cite{hartnol, S.A}, we can impose the boundary
condition that either $ J_{+} $ or $ J_{-} $ vanishes. Hereafter,
we consider the case with $ J_{+}=0 $. The case with $ J_{-}=0 $
was already studied in \cite{Roy2}, although their final
expression for critical magnetic field seems to be in error.

Now, anticipating the matching technique
\cite{gregory,zeng,Q,Barc,kanno,xiao}, first we consider the
solutions of the gauge field $ \phi(z) $ and the scalar field $
\psi(z) $, using the boundary conditions from Eqs.\ (\ref{ph}) and
(10) near the horizon ($ z = 1 $). Both $ \psi(z) $ and $ \phi(z)
$ are Taylor expanded near the horizon ($z=1$) as
\begin{eqnarray}\label{taylorphi}
\phi(z) &=& \phi(1)-\phi'(1) (1-z)+\frac{1}{2} \phi''(1) (1-z)^2+
... \nonumber \\ &\thickapprox& -\phi'(1) (1-z)+\frac{1}{2}
\phi''(1) (1-z)^2,
\end{eqnarray}
\begin{eqnarray}\label{taylorpsi}
\psi(z)=\psi(1)-\psi'(1)(1-z)+\frac{1}{2}\psi''(1)(1-z)^2+...,
\end{eqnarray}
where without loss of generality we choose $ \phi'(1)<0$ and $
\psi(1)>0 $. On the other hand, near horizon $(z = 1 )$ from Eqs.\
(\ref{phiz}) and (\ref{psiz}), and using $ f'(1)=-(d-1)r_{+}^2 $
and $ f''(1)=6r_{+}^2-(d-3)(d-4)r_{+}^2 $, we obtain
\begin{eqnarray}\label{phi1}
\phi''(1)=\bigg(\frac{d-4q}{2q-1}\bigg)\phi'(1)+\frac{\psi^2(1)
r_{+}^{2q-2}}{(-1)^{3q}2^q\beta q(2q-1)(d-1)} [\phi'(1)]^{3-2q},
\end{eqnarray}
\begin{eqnarray}\label{psi1}
\psi''(1)=\frac{m^2}{d-1}\bigg(1+\frac{m^2}{2(d-1)}\bigg)\psi(1)-\frac{\phi'^2(1)
\psi(1)}{2 r_{+}^2(d-1)^2}.
\end{eqnarray}
Substituting Eqs.\ (\ref{phi1}) and (\ref{psi1}) in Eqs.\
(\ref{taylorphi}) and (\ref{taylorpsi}), we finally obtain
\begin{eqnarray}\label{1}
\phi(z)\approx-\phi'(1)(1-z)+\frac{1}{2}\Bigg\{ \frac{d-4q}{2q-1}
+\frac{\psi^2(1)[\phi'(1)]^{2-2q} r_{+}^{2q-2}}{(-1)^{3q}2^q\beta
q (2q-1)(d-1)} \Bigg\}\phi'(1)(1-z)^2,
\end{eqnarray}
\begin{eqnarray}\label{2}
\psi(z)\approx
\left(1+\frac{m^2}{d-1}\right)\psi(1)-\frac{m^2}{d-1}z\psi(1)+\frac{1}{2}\Bigg\{\frac{m^2}{d-1}\bigg(1+\frac{m^2}{2(d-1)}\bigg)-\frac{\phi'^2(1)}{2
r_{+}^2(d-1)^2}\Bigg\} \psi(1)(1-z)^2.
\end{eqnarray}
Now by matching these two sets of asymptotic solutions at some
intermediate point $ z = z_{m} $, namely matching Eqs.\ (\ref{m})
and (\ref{J}), respectively, with Eqs.\ (\ref{1}) and (\ref{2}),
 we arrive at the following set of equations,
\begin{eqnarray}
\mu-\frac{\rho^{\frac{1}{2q-1}}}{r_{+}^{\frac{d-2}{2q-1}-1}}z_{m}^{\frac{d-2}{2q-1}-1}
=\omega(1-z_{m})-\frac{1}{2}\Bigg\{\frac{d-4q}{2q-1}+\frac{\alpha^2(-\omega)^{2-2q}
r_{+}^{2q-2}}{(-1)^{3q}2^q\beta q (2q-1)(d-1)}
\Bigg\}\omega(1-z_{m})^2,
\end{eqnarray}

\begin{eqnarray}\label{J-}
J_{-}z_{m}^{\lambda_{-}}=\bigg(1+\frac{m^2}{d-1}\bigg)\alpha-\frac{m^2}{d-1}
z_{m}\alpha+\frac{1}{2}\Bigg\{\frac{m^2}{d-1}\bigg(1+\frac{m^2}{2(d-1)}\bigg)-\frac{\omega^2}{2
r_{+}^2(d-1)^2}\Bigg\}\alpha(1-z_{m})^2,
\end{eqnarray}
where we have defined $\alpha\equiv\psi(1)$ and $
\omega\equiv-\phi'(1) $ ($ \alpha, \omega
> 0 $). Note that according to the matching method, not only
functions $\phi(z)$ and $\psi(z)$ should match at $z_m$, but their
derivatives must also match at intermediate point $z_m$. The latter
implies that we have the following two equations
\begin{eqnarray}\label{mu}
\left({\frac{d-2}{2q-1}-1}\right)\frac{\rho^{\frac{1}{2q-1}}}{r_{+}^{\frac{d-2}{2q-1}-1}}
z_{m}^{\frac{d-2}{2q-1}-2}=\omega-\Bigg\{\frac{d-4q}{2q-1}
+\frac{\alpha^2(-\omega)^{2-2q} r_{+}^{2q-2}}{(-1)^{3q}2^q\beta q
(2q-1)(d-1)} \Bigg\}\omega(1-z_{m}),
\end{eqnarray}
\begin{eqnarray}\label{J-zm}
\lambda_{-}J_{-}z_{m}^{\lambda_{-}-1}=-\frac{m^2}{d-1}\alpha-\Bigg\{\frac{m^2}
{d-1}\bigg(1+\frac{m^2}{2(d-1)}\bigg)-\frac{\omega^2}{2
r_{+}^2(d-1)^2}\Bigg\}\alpha(1-z_{m}).
\end{eqnarray}
From Eq.\ (\ref{mu}), after using Eq.\ (\ref{T}), we find
\begin{eqnarray}\label{alpha2}
\alpha^2=\frac{(-1)^{5q-1}2^{q}q \beta
(d-1)}{(1-z_{m})\tilde{\omega}^{2(1-q)}}
\left[(z_{m}-1)d-1+(6-4z_{m})q\right]\left(\frac{T_{c}}{T}\right)^{(\frac{d-2}{2q-1})}
\left[1-\bigg(\frac{T}{T_c}\bigg)^{(\frac{d-2}{2q-1})}\right],
\end{eqnarray}
where $ \tilde{\omega}={\omega}/{r_{+}} $, and  $ T_{c} $ is
obtained as
\begin{equation}\label{Tc}
T_{c}=\kappa \rho^{\frac{1}{d-2}},
\end{equation}
where
\begin{eqnarray}\label{kappa}
\kappa=\frac{(d-1)z_m^{\frac{d-4q}{d-2}}}{4\pi\tilde{\omega}^{\frac{2q-1}{d-2}}}\bigg[\frac{(d-2q-1)}
{(z_{m}-1)d-1+(6-4z_{m})q}\bigg]^{\frac{2q-1}{d-2}}.
\end{eqnarray}
From Eqs.\ (\ref{J-}) and (\ref{J-zm}), we have
\begin{eqnarray}
\tilde{\omega}=\sqrt{m^4+\frac{2(d-1)}{(1-z_{m})(2z_{m}+\lambda_{-}(1-z_{m}))}\bigg[
m^2\bigg(\lambda_{-}
(z_{m}^2-4z_{m}+3)-2z_{m}(z_{m}-2)\bigg)+2\lambda_{-}(d-1)\bigg]},
\end{eqnarray}
\begin{eqnarray}\label{J1}
J_{-}=\frac{2(d-1)+m^2(1-z_{m})}{z_{m}^{(\lambda_{-}-1)}(d-1)(2z_{m}+\lambda_{-}(1-z_{m}))}\alpha.
\end{eqnarray}
Near the critical point $ T \sim T_{c} $, after using Eq.\
(\ref{alpha2}) and the definition of distance parameter $
t\equiv1-{T}/{T_c} $, we get
\begin{equation}\label{alpha}
\alpha=\sqrt{\mathcal{A}}\sqrt{\frac{d-2}{2q-1}t},
\end{equation}
where
\begin{eqnarray}
\mathcal{A}=\frac{(-1)^{5q-1}2^{q}q \beta (d-1)}{(1-z_{m})\tilde{\omega}^{2(1-q)}} \bigg[(z_{m}-1)d-1+(6-4z_{m})q\bigg].
\end{eqnarray}
Finally, combining Eqs.\ (\ref{T}), (\ref{J1}), and (\ref{alpha})
near the critical temperature, the condensation operator can be
calculated as
\begin{eqnarray}
<\mathcal{O}_{-}>=\sqrt{2}r_{+}J_{-}=\gamma T_{c}\sqrt{1-\frac{T}{T_{c}}},
\end{eqnarray}
where
\begin{eqnarray}
\gamma=\frac{4\sqrt{2}\pi\sqrt{\mathcal{A}}}{(d-1)^2}\sqrt{\frac{d-2}{2q-1}}
\frac{2(d-1)+m^2(1-z_{m})}{{z_{m}^{(\lambda_{-}-1)}[2z_{m}+\lambda_{-}(1-z_{m})]}}.
\end{eqnarray}
\begin{center}
\begin{tabular}{|c|c|c|}
\hline
$ d $\quad &   $ q $\quad &   $\kappa $\quad  \\
\hline
$4$ \quad &   1 \quad &   0.142\quad  \\
\hline
$5$ \quad&   5/4\quad &   0.197\quad   \\
\hline
$6$ \quad &   6/4\quad &   0.231\quad   \\
\hline
$7$ \quad &   7/4 \quad &   0.259\quad  \\
\hline
$8$ \quad &   2 \quad &   0.285\quad  \\
\hline
\end{tabular}
\\[0pt]
Table $1$: The values of $ \kappa={T_c}/{\rho^{1/(d-2)}} $ for
different values of power parameter $ q={d}/{4} $, with $ z_m=0.5
$ and $ m^2=-2 $. \label{tab1}
\end{center}
We see from Table $ 1 $ that the critical temperature increases
with increasing power parameter $ q =d/4$, or dimensions $ d $, that is consistent with the general theory of critical phenomena.
There are two cases of physical interest in Table 1. The case of
$d=4$ corresponds to the $(2+1)$-dimensional holographic
superconductor in the usual Maxwell field $q=1$. The other case of
physical interest is $d=5$, which corresponds to a
$(3+1)$-dimensional holographic superconductor with conformally
invariant power-Maxwell field $q=5/4$.

\section{Effect of external magnetic field}
According to the gauge/gravity duality, the asymptotic value of
magnetic field in the bulk, corresponds to the magnetic field in
the boundary field theory: $  B(x) = F_{xy} (x, z \rightarrow 0) $
\cite{Albash,Albash2}. Considering the fact that near the upper
critical magnetic field of the continuous phase transition $
B_{\rm c2}$, the condensate order parameter is small, we can
therefore consider the scalar field $ \psi $ as a small
perturbation near $ B_{\rm c2} $. This allows us to adopt the
following ansatz for the gauge field and the scalar field
\cite{Albash,Albash2,Roy1,Roy2}
\begin{eqnarray}
A_{t}=\phi(z), \quad A_{y}=B x, \quad \psi=\psi(x,z).
\end{eqnarray}
With this, the equation for the scalar field $ \psi $ becomes
\begin{eqnarray}\label{psixz}
\psi''(x,z)&+&\bigg(\frac{f'(z)}{f(z)}-\frac{d-4}{z}\bigg)
\psi'(x,z)-\frac{m^2 r_{+}^2 \psi(x,z)}{z^4 f(z)} + \frac{r_{+}^2
\phi^2(z)\psi(x,z)}{z^4 f^2(z)}+ \frac{1}{z^2
f(z)}\left(\partial_{x}^2\psi - B^2 x^2 \psi\right)=0.
\end{eqnarray}
In order to solve Eq.\ (\ref{psixz}), we take the following separable form
\begin{equation}\label{3}
\psi(x,z)=X(x) R(z).
\end{equation}
Substituting Eq.\ (\ref{3}) into Eq.\ (\ref{psixz}), we arrive at
\begin{eqnarray}
z^2
f(z)\bigg[\frac{R''}{R}+\bigg(\frac{f'(z)}{f(z)}-\frac{d-4}{z}\bigg)\frac{R'}{R}+\frac{r_{+}^2\phi^2}{z^4f^2}-\frac{m^2
r_{+}^2}{z^4 f}\bigg]-\bigg[-\frac{X''}{X}+B^2x^2\bigg]=0.
\end{eqnarray}
$ X (x) $ is governed by the equation of a harmonic
oscillator with frequency determined by $ B $:
\begin{eqnarray}
-X''(x)+B^2x^2X(x)=\lambda_{n} B X(x),
\end{eqnarray}
where $ \lambda_n= 2n + 1  $. For $n=0$, $ R(z) $ satisfies
\begin{eqnarray}\label{Rz}
R''(z)+\bigg(\frac{f'(z)}{f(z)}-\frac{d-4}{z}\bigg) R'(z)-\frac{m^2 r_{+}^2 R(z)}
{z^4 f(z)}+\frac{r_{+}^2\phi^2(z)R(z)}{z^4f^2(z)}=\frac{BR(z)}{z^2 f(z)}.
\end{eqnarray}
Now at the horizon $ z = 1 $, and using Eq.\ (\ref{Rz}),
$ \phi(1)=0 $, $ f'(1)=-(d-1)r_{+}^2 $, and $ f''(1)=6r_{+}^2-(d-3)(d-4)r_{+}^2 $, we obtain
\begin{equation}\label{R'}
R'(1)=-\bigg(\frac{B}{(d-1)r_{+}^2}+\frac{m^2}{d-1}\bigg)R(1),
\end{equation}
\begin{eqnarray}\label{R''}
R''(1)=\Bigg\{\frac{m^2}{d-1}\bigg(1+\frac{B}{(d-1)
r_{+}^2}+\frac{m^2}{2(d-1)}\bigg)+\frac{B^2}{2(d-1)^2
r_{+}^4}-\frac{\phi'^2(1)}{2(d-1)^2 r_{+}^2}\Bigg\}R(1).
\end{eqnarray}
In the asymptotic region ($ z\rightarrow 0 $), the solution behaves as
\begin{equation}\label{5}
R(z)=J_{-}z^{\lambda_{-}}+J_{+}z^{\lambda_{+}}.
\end{equation}
In the presence of external magnetic field, we use the matching method and the Taylor expansion of $ R(z) $
near the horizon
\begin{eqnarray}\label{R(z)}
R(z)=R(1)-R'(1)(1-z)+\frac{1}{2}R''(1)(1-z)^2+ \cdots\ .
\end{eqnarray}
Substituting Eqs. (\ref{R'}) and (\ref{R''}) in Eq.\ (\ref{R(z)}), we have
\begin{eqnarray}\label{4}
R(z)&\thickapprox&
R(1)+(1-z)\bigg(\frac{B}{(d-1)r_{+}^2}+\frac{m^2}{d-1}\bigg)R(1)
\nonumber\\&&+\frac{1}{2}
\bigg[\frac{m^2}{d-1}\bigg(1+\frac{B}{(d-1) r_{+}^2}
+\frac{m^2}{2(d-1)}\bigg)+\frac{B^2}{2(d-1)^2
r_{+}^4}-\frac{\phi'^2(1)}{2(d-1)^2 r_{+}^2}\bigg]R(1)(1-z)^2 .
\end{eqnarray}
The two solutions given in Eq.\ (\ref{5}) and Eq.\ (\ref{4}) are
connected smoothly at an intermediate point $ z_m $. Thus, we find
that
\begin{eqnarray}
J_{-}z_{m}^{\lambda_{-}}&=&
R(1)+(1-z_{m})\bigg(\frac{B}{(d-1)r_{+}^2}+\frac{m^2}{d-1}\bigg)R(1)
\nonumber+\frac{1}{2} \Bigg\{\frac{m^2}{d-1}\bigg(1+\frac{B}{(d-1)
r_{+}^2}+\frac{m^2}{2(d-1)}\bigg)\\&&+\frac{B^2}{2(d-1)^2
r_{+}^4}-\frac{\phi'^2(1)}{2(d-1)^2 r_{+}^2}\Bigg\}
R(1)(1-z_{m})^2,
\end{eqnarray}
\begin{eqnarray}
\lambda_{-}J_{-}z_{m}^{\lambda_{-}-1}&=&-\bigg(\frac{B}{(d-1)r_{+}^2}+\frac{m^2}{d-1}\bigg)R(1)
\nonumber- \Bigg\{\frac{m^2}{d-1}\bigg(1+\frac{B}{(d-1)
r_{+}^2}+\frac{m^2}{2(d-1)}\bigg)\\&&+\frac{B^2}{2(d-1)^2
r_{+}^4}-\frac{\phi'^2(1)}{2(d-1)^2 r_{+}^2}\Bigg\}R(1)(1-z_{m}).
\end{eqnarray}
From the above equations, we get
\begin{eqnarray} \label{B2}
B^2+2nr_{+}^2B+pr_{+}^4-\phi'^2(1)r_{+}^2=0,
\end{eqnarray}
where
\begin{eqnarray}
n=m^2+2(d-1)\Bigg\{\frac{\lambda_{-}
(1-z_{m})+z_{m}}{(1-z_{m})\left[(2-\lambda_{-})z_{m}+\lambda_{-}\right]}\Bigg\},
\end{eqnarray}
\begin{eqnarray}
p=m^4+\frac{2(d-1)\left[2\lambda_{-}(d-1)+m^2\left((3-4z_{m}+z_{m}^2)\lambda_{-}-2z_{m}(z_{m}-2)\right)\right]}{(1-z_{m})
\left[(2-\lambda_{-})z_{m}+\lambda_{-}\right]}.
\end{eqnarray}
It is easy to see that Eq. (\ref{B2}) has a solution
\begin{equation}\label{B}
B=\sqrt{\phi'^2(1)r_{+}^2+r_{+}^4(n^2-p)}-nr_{+}^2.
\end{equation}
Now consider the case for which the value of the external magnetic
field is very close to the upper critical value, i.e.\ $ B
\sim B_{\rm c2} $. This implies a vanishingly small condensate, and
therefore we can ignore all the quadratic terms in $ \psi $. With
this approximation, from Eq.\ (\ref{phiz}) we obtain
\begin{eqnarray}
\phi''(z)-\frac{1}{z}\bigg(\frac{d-2}{2q-1}-2\bigg)\phi'(z)=0.
\end{eqnarray}
From the above equation and using Eq.\ (\ref{m}), we arrive at
\begin{eqnarray}\label{phi}
\phi'^2(1)r_{+}^2=\frac{\rho^{\frac{2}{2q-1}}}{r_{+}^{\frac{2(d-2)}{2q-1}}}r_{+}^4 \bigg(\frac{d-2}{2q-1}-1\bigg)^2.
\end{eqnarray}
From Eqs.\ (\ref{Tc}) and (\ref{kappa}), we have
\begin{eqnarray}\label{rho}
\rho=\frac{\left[4\pi T_{c}(0)\right]^{(d-2)}
\tilde{\omega}^{2q-1}}
{z_{m}^{(d-4q)}(d-1)^{(d-2)}\bigg[\frac{d-2q-1}{(z_{m}-1)d-1+(6-4z_{m})q}\bigg]^{2q-1}}.
\end{eqnarray}
On substituting Eqs.\ (\ref{phi}) and (\ref{rho}) into Eq.\ (\ref{B}), we finally obtain
\begin{eqnarray}\label{Bc}
\frac{B_{\rm
c2}(T)}{T_{c}^2(0)}&=&\frac{16\pi^2}{(d-1)^2}\left(\frac{T}{T_{c}(0)}\right)^{\frac{4q-d}{2q-1}}
\nonumber
\Bigg\{\sqrt{\frac{\tilde{\omega}^2\left[(z_{m}-1)d-1+(6-4z_{m})q\right]^2}{(2q-1)^2z_{m}^{\frac{2(d-4q)}
{2q-1}}}+(n^2-p)\left(\frac{T}{T_{c}(0)}\right)^{\frac{2d-4}{2q-1}}}
\\&&-n\left(\frac{T}{T_{c}(0)}\right)^{\frac{d-2}{2q-1}}\Bigg\}.
\end{eqnarray}
This result makes clear the dependence of the upper critical field
on the temperature and the power parameter of the power-Maxwell
electrodynamics. The phase diagram given by Eq.\ (\ref{Bc}) is
plotted in Figs.\ 1 and 2 for different values of dimensionality $
d $ (or power parameter $q=d/4$) and different $m^2$,
respectively. It is clear from Figs.\ 1 and 2 that, consistent
with the phenomenology and the Ginzburg-Landau theory of
superconductivity, $B_{\rm c2}(T)$ vanishes linearly as
$T\rightarrow T_c(0)$. Figure 3 shows the behavior of $B_{\rm c2}$
as a function of $T$ for different values of $T_c(0)$, i.e.\ the
transition temperature in absence of magnetic field. It is evident
from Fig.\ 3 that $B_{\rm c2}(0)$ increases with $T_c(0)$, which
is again consistent with the phenomenology of superconductivity.
It is worth noting that the reasonable behavior of $B_{\rm c2}$ in
Figs.\ 1-3, which is consistent with the phenomenology  of
superconductors, is observed provided we take $q=d/4$. For example
in $d=5$ dimensions that corresponds to a (3+1)-dimensional
holographic superconductor, we should take $q=5/4$ in order to
have reasonable $B_{\rm c2}$ in the range of $T\leq T_c(0)$. This
implies that among all power-Maxwell theories with various
parameters $q$, those which are conformally invariant lead to the
magnetic field behavior of physical interest for holographic
superconductors. This may be understood as follows. For the
conformally invariant power-Maxwell theory, not only the
energy-momentum tensor is traceless, but also the electromagnetic
fields in higher dimensions have the same behavior as in the (well
established) case of (2+1)-dimensional holographic superconductor
dual of 4-dimensional gravity \cite{SheyPM}. For example, for the
usual Maxwell field in $d$ dimensions, the electric field of
charged point-like particle behaves as $E(r)\sim 1/r^{d-2}$, while
for the conformally invariant power-Maxwell field in
$d$-dimensional spacetime, the electric field of point charge is
independent of the dimensionality and varies as $E(r)\sim
1/r^{2}$, exactly like in four dimensions \cite{SheyPM}. In other
words, for conformally invariant Maxwell field, the magnetic field
in the $d$-dimensional bulk and on its $(d-1)$-dimensional
boundary has the same behavior as in the usual holographic
superconductor in $d=4$ spacetime.
\begin{figure}
\centering\includegraphics[scale=.5]{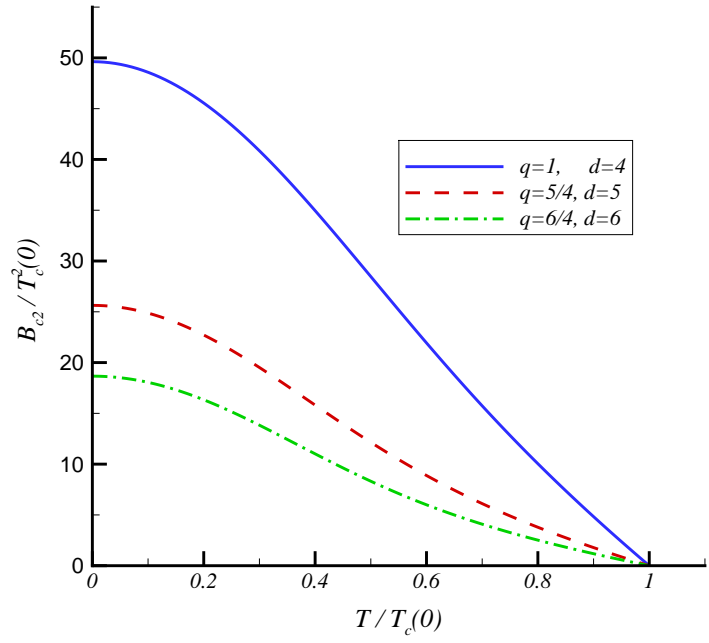}\caption{The behavior of
$ B_{\rm c2}/T_{c}^2(0)$ in terms of $T/T_{c}(0)$ for different
values of $q$ and $ d $, with $ z_m=0.5 $ and $ m^2=-2
$.}\label{fig1}
\end{figure}
\begin{figure}
\centering\includegraphics[scale=.5]{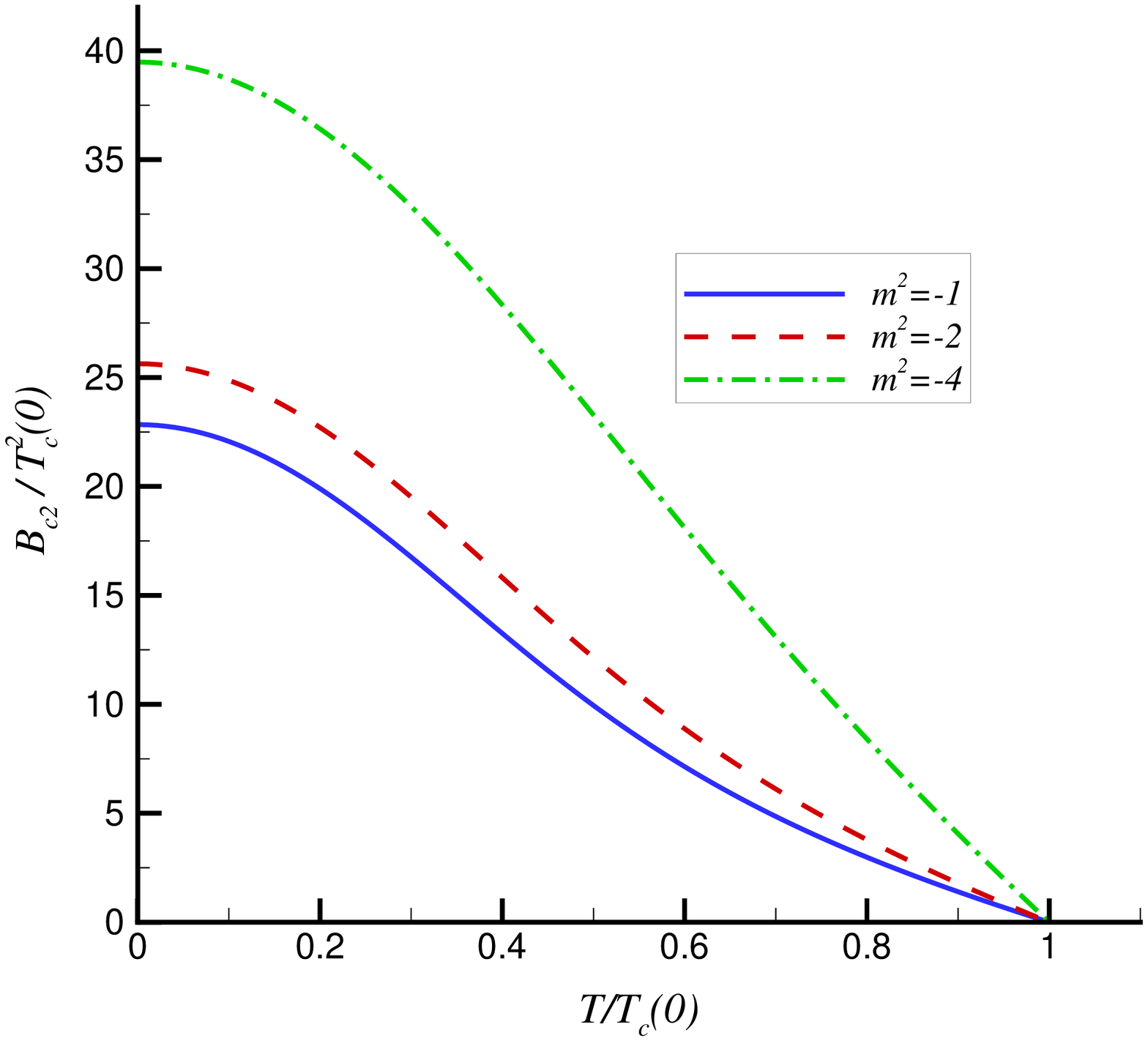}\caption{The behavior of
$ B_{\rm c2}/T_{c}^2(0)$ in terms of $T/T_{c}(0)$ for different
values of $m^2$ and $ z_m=0.5 $, with $ d=5$ and $
q=5/4$.}\label{fig2}
\end{figure}
\begin{figure}
\centering\includegraphics[scale=.5]{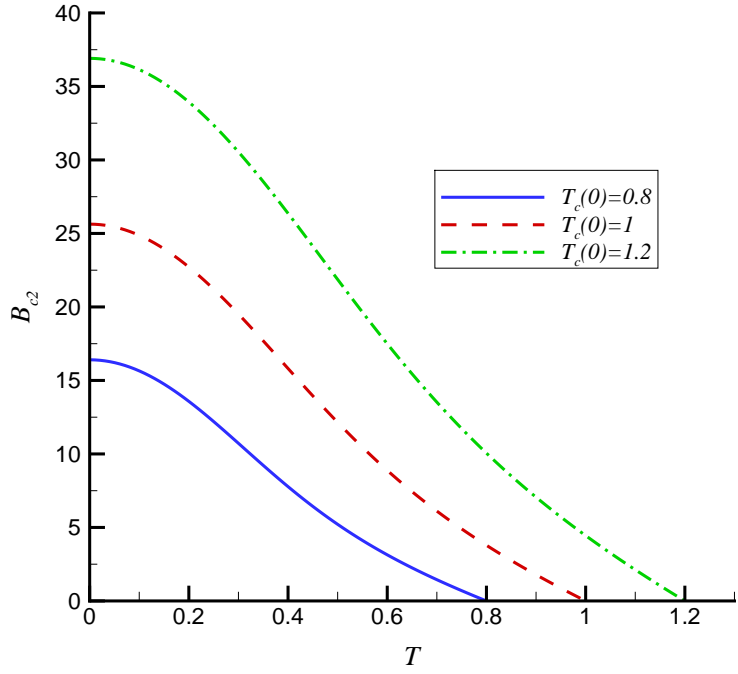}\caption{The upper
critical field $ B_{c2}$ as a function of temperature $T $ for
different values of $ T_c(0) $, with $ z_m=0.5 $, $ d=5 $, $ q=5/4
$, and $ m^2=-2 $.}\label{fig3}
\end{figure}
\begin{figure}
\centering\includegraphics[scale=.5]{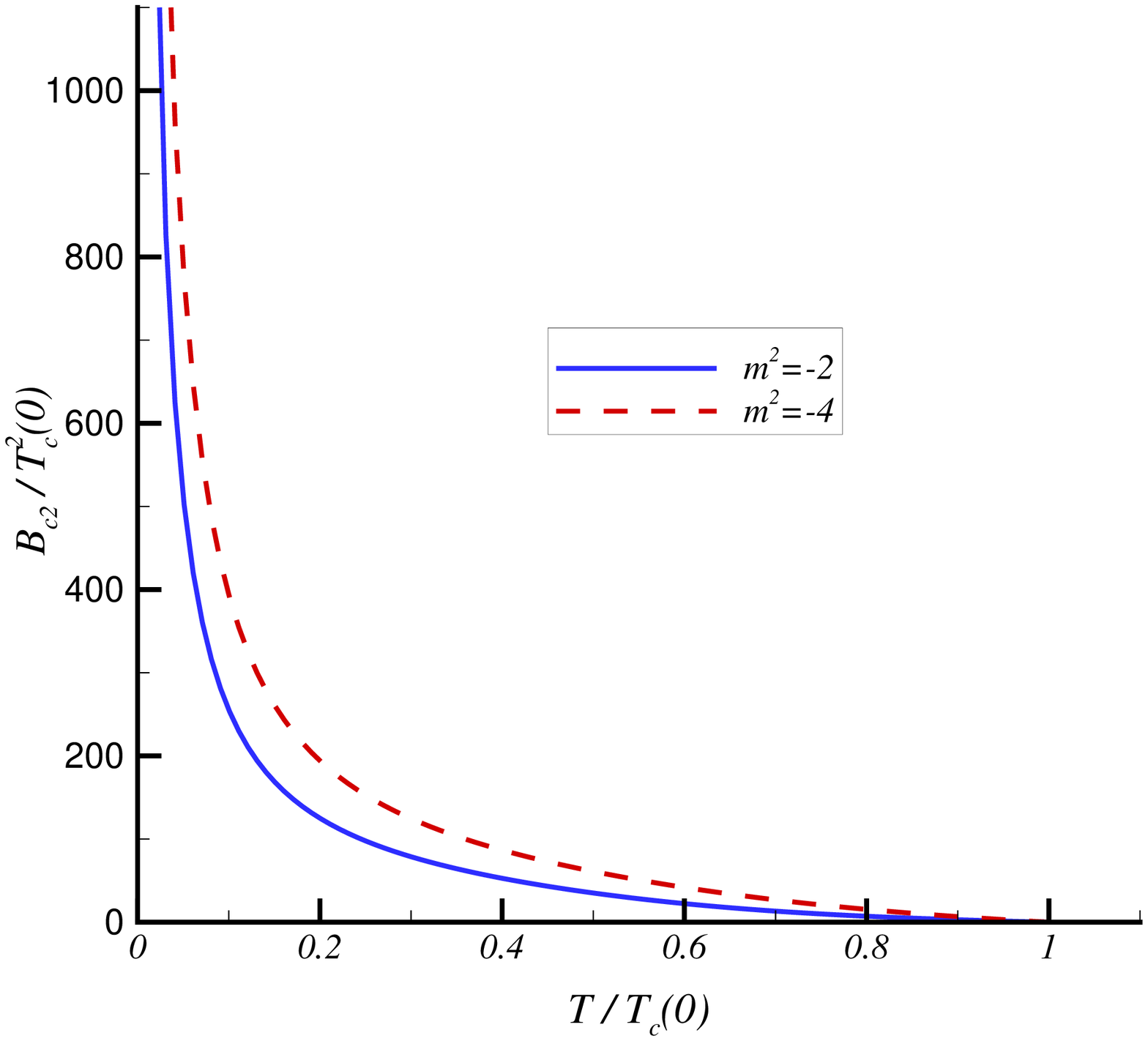}\caption{The behavior of
$ B_{\rm c2}/T_{c}^2(0)$ in terms of $T/T_{c}(0)$ for different
values of $m^2$ and $ z_m=0.5 $, with  $ d=5 $ and $q=1
$.}\label{fig4}
\end{figure}
\begin{figure}
\centering\includegraphics[scale=.5]{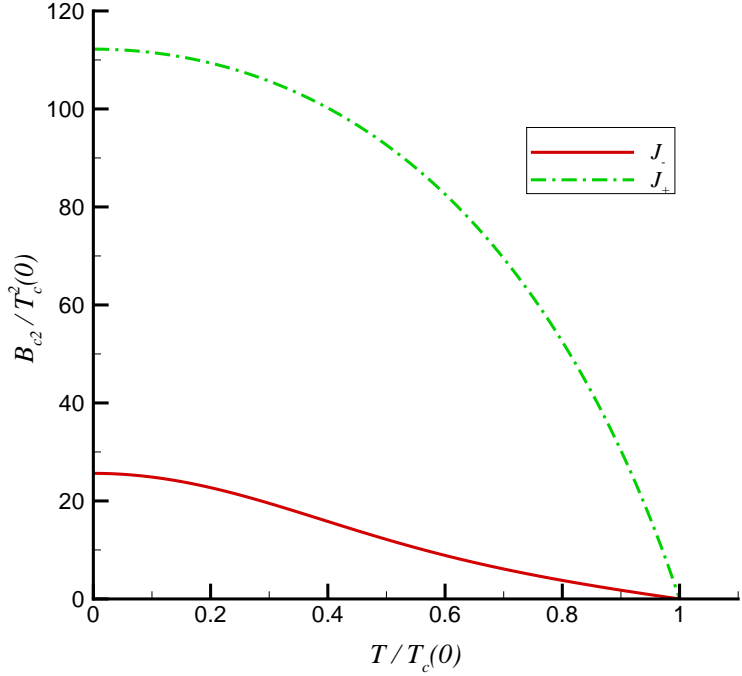}\caption{The behavior of
$ B_{c2}/T_{c}^2(0)$ in terms of $T/T_{c}(0) $ for different
non-zero boundary conditions $ J_{-} $ and $ J_{+} $, with $
z_m=0.5 $, $ d=5 $, $ q=5/4 $, and $ m^2=-2 $ .}\label{fig5}
\end{figure}

In order to clarify the above argument, let us have a closer look
at expression (\ref{Bc}). For conformally invariant case
$q=d/4$, the critical magnetic field in $d$-dimensions reduces to
\begin{eqnarray} \label{Bc2}
\frac{B_{\rm c2}(T)}{T_{c}^2(0)}=\frac{16\pi^2}{(d-1)^2}
\bigg[\sqrt{\tilde{\omega}^2+(n^2-p)\bigg(\frac{T}{T_{c}(0)}\bigg)^{4}}-n\bigg(\frac{T}{T_{c}(0)}\bigg)^{2}\bigg].
\end{eqnarray}
It is important to note that in this case the functional
dependence of $B_{\rm c2}(T)$ is the same as that of four-dimensional
Maxwell case. This is an expected result since for conforamally invariant Maxwell field, the magnetic field in
all dimensions has the same behavior as in four dimensions. For
$d=5$, expression (\ref{Bc2}) becomes
\begin{eqnarray} \label{Bc3}
\frac{B_{\rm c2}(T)}{T_{c}^2(0)}=\pi^2
\Bigg\{\sqrt{\tilde{\omega}^2+(n^2-p)\bigg(\frac{T}{T_{c}(0)}\bigg)^{4}}-n\bigg(\frac{T}{T_{c}(0)}\bigg)^{2}\Bigg\}.
\end{eqnarray}
On the other hand if we consider the Maxwell case ($q=1$) in five
dimensions $(d=5)$, then  Eq. (\ref{Bc}) becomes
\begin{eqnarray} \label{Bc4}
\frac{B_{\rm
c2}(T)}{T_{c}^2(0)}=\pi^2\left(\frac{T}{T_{c}(0)}\right)^{-1}
\Bigg\{\sqrt{\tilde{\omega}^2+(n^2-p)\bigg(\frac{T}{T_{c}(0)}\bigg)^{6}}-n\bigg(\frac{T}{T_{c}(0)}\bigg)^{3}\Bigg\}.
\end{eqnarray}
The main difference between expressions of critical magnetic field
given in Eq.\ (\ref{Bc3}) for the conformally invariant Maxwell field
($ d=5,\, q=5/4$) and the usual Maxwell field ($d=5,\,q=1$) given by
Eq.\ (\ref{Bc4}), is the appearance of the
$\left[{T}/{T_{c}(0)}\right]^{-1}$ term in the latter case.
Clearly this term diverges as
$\left[{T}/{T_{c}(0)}\right]\rightarrow0$. The behavior of
$B_{\rm c2}$  in terms of ${T}/{T_{c}}(0)$  for $q=1$ and $d=5$  is
shown in Fig.\ 4.  From this figure, it is obvious that the critical
magnetic field diverges as  $T\rightarrow 0$. This behavior is
unacceptable and as we mentioned only the conformally invariant
Lagrangian with $q=d/4$ leads to the desired form of critical magnetic
field consistent with the phenomenology of superconductors. Finally,
Fig.\ 5 shows the phase diagrams obtained by imposing the
different boundary conditions of non-zero $ J_- $, as expressed by
Eq.\ (\ref{Bc}), and non-zero $ J_+ $, also considered in Ref.\ \cite{Roy2}.
\section{Conclusions}
In this paper, the properties of ($d-1$)-dimensional $s$-wave
holographic superconductor in the presence of conformally
invariant power-Maxwell correction to the usual Einstein-Maxwell
action, were investigated. It must be noted that the power-Maxwell Lagrangian
in $d$ dimensions is invariant under
conformal transformation, i.e.\ $ g_{\mu\nu} \rightarrow
\Omega^2g_{\mu\nu}$ and $A_\mu \rightarrow A_\mu $, provided we
take the power parameter as $ q= d/4 $.

In the absence of external magnetic field, we have found the
critical temperature $T_c$ to vary as $ \rho^{{1}/{(d-2)}}$ for
all values of the power parameter $q$, with a proportionality
constant $\kappa$ that increases with $q$ or the dimensionality
$d$, as expected. The variation of order parameter with the
temperature is found to be $ <\mathcal{O}>  \propto
\sqrt{1-\frac{T}{T_c}} $ with a critical exponent $\beta={1}/{2}
$, which is characteristic of systems with mean field behavior,
the prime examples of which are the superconductors.

Furthermore, an analytic investigation of the effects of an
external magnetic field was made by employing the matching
technique. We considered the probe limit in which the scalar and
gauge fields do not affect the background metric.  The phase
behavior of the ($d-1$)-dimensional holographic superconductor was
obtained by computing the upper critical magnetic field in the
presence of conformally invariant power-Maxwell
electrodynamics.  It was found that consistent with the
phenomenology and Ginzburg-Landau theory of superconductivity, the
upper critical field $B_{\rm c2}(T)$ vanishes linearly as
$T\rightarrow T_c(0)$. Also, it became evident that $B_{\rm
c2}(0)$ increases with $T_c(0)$, which is again consistent
with the phenomenology of superconductors. The case of physical interest
in $d=5$ spacetime dimensions, corresponding to (3+1)-dimensional holographic
superconductor, was considered in detail with both the conformally invariant $q=5/4$ and the usual power-Maxwell parameter $q=1$, and plotted in the
figures. Interestingly enough, we observed that in the
presence of power-Maxwell electrodynamics, the critical
magnetic field $B_{\rm c2}$ has reasonable behavior consistent
with the phenomenology of superconductors provided we consider the
conformally invariant case, i.e.\ $q=d/4$. This may be due to the
fact that for the conformally invariant power-Maxwell theory, the
electromagnetic fields in higher dimensions have the same behavior
as in the four-dimensional gravity dual of the well established (2+1)-dimensional holographic superconductor \cite{SheyPM}.
In particular, we found that
for  $q=1$ and $d=5$, the critical magnetic field $B_{\rm c2}$ diverges
as $T\rightarrow 0$. This behavior is physically  unacceptable and comes
from the fact that in the case $q=1$ and $d=5$, the power-Maxwell
Lagrangian is not conformally invariant. This point, however,
deserves further investigation.

\acknowledgments{We thank the Research Council of Shiraz
University. This work has been supported financially by Research
Institute for Astronomy and Astrophysics of Maragha (RIAAM),
Iran.}

\end{document}